\documentclass[%
 aip,
 amsmath,amssymb,
 reprint,%
]{revtex4-1}

\usepackage{color}
\usepackage{graphicx}% Include figure files
\usepackage{dcolumn}% Align table columns on decimal point
\usepackage{bm}% bold math
\usepackage[utf8]{inputenc}
\usepackage[T1]{fontenc}
\usepackage{mathptmx}
\usepackage{etoolbox}

\newcommand{\sig}{\boldsymbol{\hat{\sigma}}}
\providecommand{\abs}[1]{\lvert#1\rvert}

\makeatletter
\def\@email#1#2{%
 \endgroup
 \patchcmd{\titleblock@produce}
  {\frontmatter@RRAPformat}
  {\frontmatter@RRAPformat{\produce@RRAP{*#1\href{mailto:#2}{#2}}}\frontmatter@RRAPformat}
  {}{}
}%
\makeatother
\begin{document}

\preprint{AIP/123-QED}

\title[]{Dynamics of an inelastic tagged particle under strong confinement}
% Force line breaks with \\
\author{P. Maynar}\email{maynar@us.es}
\author{M. I. Garc\'ia de Soria}
\author{J. J. Brey}
\affiliation{F\'isica Te\'orica, Universidad de Sevilla, Apartado de
  Correos 1065, E-41080, Sevilla, Spain}%
\affiliation{Institute for Theoretical and Computational
  Physics. Facultad de Ciencias. Universidad de Granada, E-18071,
  Granada, Spain%\\This line break forced% with \\
}%

\date{\today}% It is always \today, today,
             %  but any date may be explicitly specified

\begin{abstract}
The dynamics of a tagged particle immersed in a fluid of particles of
the same size but different mass is studied when the system is
confined between two hard parallel 
plates separated a distance smaller than twice the diameter of the
particles. %(in such a way that the system is
%quasi-two-dimensional). 
The collisions between particles are
inelastic while the collisions of the particles with the hard walls
inject energy in the direction perpendicular to the wall, so that
stationary states can be reached in the long-time limit. The velocity
distribution of the tagged particle verifies a Boltzmann-Lorentz-like
equation that is solved assuming that it is a spatially homogeneous
gaussian distribution with two different temperatures (one
associated to the motion parallel to the wall and another associated to the
perpendicular direction). It is found that 
the temperature perpendicular to the wall diverges when the tagged
particle mass approaches a critical mass from below, while the
parallel temperature remains finite. Molecular Dynamics simulation results
agree very well with the theoretical predictions for tagged particle
masses below the critical mass. The measurements of the velocity
distribution function of the tagged particle confirm that it is
gaussian if the mass is not close to the critical mass, while it
deviates from gaussianity when approaching the critical mass. Above
the critical mass, the velocity distribution function is very far from
a gaussian, being the marginal distribution in the perpendicular
direction bimodal and with a much larger variance than the
one in the parallel direction. 
\end{abstract}

\maketitle

\section{Introduction}

Granular systems are ensembles of macroscopic particles whose
interactions are dissipative, in the sense that when two particles
(grains) collide, part of the center of mass kinetic energy is
transferred to internal degrees of freedom. In the fluid regime, the dynamics of the system 
(understood as the ensemble of grains independently of their internal
structure) can be thought as a sequence of inelastic binary
collisions and the system is reminiscent to a molecular gas. A kinetic
theory description is applicable in this case \cite{gs95} and, at a
larger scale, hydrodynamics has been shown to
describe the macroscopic behavior of the system in many situations, as
well as to explain 
several instabilities that appears in different contexts \cite{bdks98,
  g03, at06}. 

When energy is continuously supplied to the system, stationary states can be reached
in which the energy injected is compensated by the energy dissipated in
collisions. The energy injection mechanism can be very simple, for
example, by just  agitating the box in which the system is, or
by vibrating one of the confining walls. 
Typically, non-homogeneous stationary states are obtained as it
can be seen from the hydrodynamic equations \cite{bdks98}. Nevertheless,
if the system is agitated vertically and its height is small (of
the order of the particles diameter), in such a way that it is
quasi-two-dimensional (Q2D), stationary 
states are reached that can be considered spatially homogeneous
%because they are spatially homogeneous in the horizontal direction
%and the gradients in the vertical direction can be neglected. 
\cite{ou98, peu02, rcbhs11}. These
configurations are specially interesting because, as granular
systems are intrinsically out of equilibrium, the generated non-equilibrium
homogeneous stationary state can be used to test experimentally many of the out of
equilibrium statistical
mechanics machinery in a very simple situation. In fact, 
in the last two decades, a lot of experiments have been performed
exploiting the above mentioned property \cite{ou98, peu02, rcbhs11,
  crbhs12, mvprkeu05, 
  ou05, gsvp11, pggs12, cms12, gs18}. Devices with or without a top
lib have been used, being gravity the
responsible of the Q2D confinement in the latter case. It is found that, for a
wide range of the parameters describing the state of the system, 
homogeneous stationary states are reached. Nevertheless, by increasing
the averaged density or by varying
some of the parameters that describe the vibration of the walls, the
homogeneous state becomes unstable. Another stationary 
state is reached in which a dense aggregate, surrounded by a
more dilute hotter phase, appears. Let us also mention that, depending on the
averaged density, the coexistence is between a solid-like phase and a
liquid-like phase \cite{ou98} or between a liquid and a gas \cite{rcbhs11}. 

Several models have been proposed in order to explain the above
mentioned instability. Particularly interesting is the one introduced
in \cite{brs13} in which the system is modeled
as an ensemble of hard disks with a collision rule modified in
such a way that, depending on the relative velocity,  energy can be
gained or lost in a collision. This model has been
widely studied finding that the homogeneous stationary state is always
stable \cite{bgm13, srb14, bmgb14, bgmb14, bbmg15, bbgm16}, so that it
can not describe the phenomenology seen 
in the experiments. In order to describe the latter, it appears
essential to take into account that energy is injected in 
the vertical direction and that it is transferred to the horizontal
degrees of freedom through inelastic collisions. Although some models
have been introduced to incorporate this ingredient \cite{rsg18}, it
seems that the simplest model is an ensemble of inelastic spheres
confined between two hard walls 
and injecting energy through the walls by some mechanism. In Refs. 
\cite{mgb19, mgb19B} this model was studied assuming that the height of the
system is smaller than twice the diameter of the particles (in order
to be Q2D), and that the bottom wall is a vibrating elastic sawtooth
wall. The top one an elastic wall at rest. 
It was shown that, for low densities, the pressure in the 
horizontal direction decays monotonically with the density (apparent negative
compressibility) triggering the instability when its horizontal
dimension is large enough (otherwise it is killed by heat
diffusion). This is in agreement with the explanation proposed in
\cite{brs13}. 

In a mixture of two species of grains of equal size but
different mass, other new instabilities have been observed. Particularly
relevant is the one studied in \cite{rpgrscm11,
  rcrs12}. Spontaneous segregation shows up with a cluster of heavier
particles surrounded by lighter ones. It is found that,  
when the system is partly segregated, there are sudden peaks of the
horizontal kinetic energy of the heavy particles (otherwise
small), that partially destroy the cluster.  
In this paper, we study a mixture of two species of grains of the same
size, but in the
simpler situation in which there is only one particle of a different
mass (the intruder). It is assumed that the bath is dilute and it is always in the
homogeneous steady state. In the same lines as in Refs. \cite{mgb19, mgb19B}, the
simplest model is considered (neglecting gravity, friction between
grains and also friction between grain and the two
  walls) but, in this case, we will 
assume that the two walls vibrate. The reason is that, in real
experiments, when the system is agitated vertically, both walls inject
energy into the system. The objective is to study the dynamics and the 
stationary states that the tagged particle eventually reaches in the long time
limit. Preliminary Molecular Dynamics (MD) results \cite{VicentePhD} have shown that, if the mass of
the intruder is close to the one of the bath particles, its 
distribution function is close to a two-temperatures gaussian, being the
horizontal and vertical temperatures of the order of the two bath
temperatures. Remarkably, for the parameters considered in \cite{VicentePhD}, 
it was also shown that, when the tagged
particle mass was only twice the one of the bath, the vertical
temperature was order of magnitudes larger. In addition, the
distribution function was not a gaussian anymore. %being even bimodal in
%the vertical direction. 
In this paper we will study more deeply these effects by kinetic
theory. More precisely, using the same arguments to the ones used to
derive the Boltzmann equation for ultra-confined hard spheres
\cite{bmg16, bgm17}, a Boltzmann-Lorentz equation that describes the
dynamics of the tagged particle is formulated. In the stationary
state, this equation is
approximately solved using a two-temperature gaussian ansatz,
finding that the vertical temperature diverges for some ``critical''
value of the tagged particle mass. This critical mass depends on the
inelasticity of the particles and on the height of the box. Moreover,
if the mass of the intruder is not close to its critical value and it
is also smaller than it, MD simulation results show that the gaussian
ansatz is a 
good approximation and a very good agreement with the theoretical
prediction is found. This agreement is progressively broken when the mass
of the intruder increases and the corresponding  
critical value is approached (above the critical mass, the
distribution function is not gaussian anymore). 

The paper is organized as follows: in the following section, the
model to be considered is introduced and the Boltzmann-Lorentz equation
describing the dynamics of the intruder is formulated. In
Sec. \ref{homoStates}, the dynamics of the intruder is studied assuming
that its one-particle distribution function is a two-temperatures
gaussian. The properties of the stationary state are also discussed. 
The theoretical predictions are compared with MD
simulation results in Sec. \ref{secSR}, and a good agreement is found in the
region of the parameters where the gaussian approximation is
fulfilled. The final section of the paper contains a short summary of
the results, whose relevance is discussed. Some details of the calculations are
presented in the Appendix.

\section{The model}\label{secModel}

The system we consider consists of an ensemble of $N$ smooth inelastic hard
spheres of mass $m$ and diameter $\sigma$, plus another inelastic particle of
mass $M$ and the same diameter. Particles are confined between two
parallel square-shaped plates of area $A$, separated a distance $h$. It
is $h<2\sigma$, so that particles can not jump over each other and the
system can be considered to be Q2D. The collision rule between the
particle of mass $M$ and the ones of the bath is
\begin{eqnarray}
\mathbf{v}'\equiv
  b_{\sig}\mathbf{v}=\mathbf{v}+\frac{m}{m+M}(1+\alpha)(\mathbf{g}\cdot\sig)\sig, 
\label{colRule1}\\
\mathbf{v}_1'\equiv
  b_{\sig}\mathbf{v}_1=\mathbf{v}_1-\frac{M}{m+M}(1+\alpha)(\mathbf{g}\cdot\sig)\sig, 
\label{colRule2}
\end{eqnarray}
where $\mathbf{v}$ and $\mathbf{v}_1$ are the velocities of the
particles of mass $M$ and $m$ respectively before the collision, 
$\mathbf{g}\equiv\mathbf{v}_1-\mathbf{v}$, $\sig$ an unit
vector joining the two particles at contact from the particle
of mass $M$ to the other one, and $\alpha$ the coefficient of normal
restitution that will be considered as constant (independent of the
relative velocity). It fulfills $0<\alpha\le 1$, being $\alpha=1$ the
elastic collision limit. We have also introduced the operator
$b_{\sig}$ that transforms the velocities of the 
particles into the respective velocities after the collision. The
collision rule for the particles of the bath is similar, taking
$M=m$ and substituting $\alpha$ by the coefficient of normal
restitution of the bath particles, $\alpha_1$. Periodic boundary
conditions are used in the horizontal directions. The bottom and top 
walls are located at $z=0$ and $z=h$ respectively
and are sawtooth type, i.e. when a
particle collides with the bottom (top) wall, the particle always
``sees'' the wall moving upwards (downwards) with velocity $v_0$ and
undergoes an elastic collision. By
introducing the unitary vectors in the direction of the axes 
$\{\mathbf{e}_x, \mathbf{e}_y, \mathbf{e}_z\}$, the particle-wall collision rules are
\begin{eqnarray}
\mathbf{v}\longrightarrow b_b\mathbf{v}\equiv v_x\mathbf{e}_x+v_y\mathbf{e}_y+(2v_0-v_z)
  \mathbf{e}_z, \\
\mathbf{v}\longrightarrow b_t\mathbf{v}\equiv v_x\mathbf{e}_x+v_y\mathbf{e}_y-(2v_0+v_z)
  \mathbf{e}_z, 
\end{eqnarray}
for the bottom and top wall respectively. We have also introduced the
corresponding operators $b_b$ and $b_t$. Note that this kind of
collisions always injects energy into the system and conserve momentum
in the direction parallel to the plates. Since momentum is conserved
in the collisions between particles, total horizontal momentum is
a constant of the motion. 

Due to the inelasticity of the particle collisions,
stationary states in which the energy lost in
collisions is compensated by the energy injected through the
walls can be obtained. The stationary states reached by the bath (the actual system
\emph{without} the particle of mass $M$) were studied in \cite{mgb19,
  mgb19B}, finding that, if the width of the system is small enough, a
spatially homogeneous stationary state is reached (in the low density
limit the gradients in the vertical direction can be neglected). In
the following, we will assume that this is the case. In
\cite{mgb19} it was shown that the distribution function, $f_1$, can
be accurately approximated by a two-temperature gaussian
\begin{equation}\label{f1maintext}
f_1(\mathbf{v})=\frac{n_1}{\pi^{3/2}w^2w_z^2}e^{-\frac{v_x^2+v_y^2}{w^2}-\frac{v_z^2}{w_z^2}}, 
\end{equation}
where $n_1=\frac{N}{(h-\sigma)A}$ is the three-dimensional density of
the gas. The thermal velocities, $w$ and $w_z$, are related to the
horizontal, $T_1$, and vertical, $T_{1,z}$, temperatures through
\begin{eqnarray}
\frac{m}{2}w^2=T_1, \quad \frac{m}{2}w_z^2=T_{1,z}.  
\end{eqnarray}
The horizontal and vertical temperatures are defined as usual in
kinetic theory
\begin{equation}\label{defTTempsBa}
n_1T_1=\frac{m}{2}\int d\mathbf{v}(v_x^2+v_y^2)f_1(\mathbf{v}), \quad
n_1T_{1,z}=m\int d\mathbf{v}v_z^2f_1(\mathbf{v}). 
\end{equation}
The explicit expressions for the steady partial temperatures in terms of the
parameters of the bath are \cite{mgb19}
\begin{equation}\label{gamma1}
\gamma_1\equiv\frac{T_{1,z}}{T_1}=\frac{12(1-\alpha_1)+(5\alpha_1-1)\epsilon^2}{(3\alpha_1+1)\epsilon^2}, 
\end{equation}
and 
\begin{equation}\label{Tbath}
T_1=\left[\frac{6\gamma_1}{\sqrt{\pi}(1+\alpha_1)\left(\gamma_1-\frac{1+\alpha_1}{2}\right)\epsilon^3\tilde{n_1}\sigma^2}\right]^2mv_0^2, 
\end{equation}
where the dimensionless height,
$\epsilon\equiv\frac{h-\sigma}{\sigma}$, and the effective
two-dimensional density, $\tilde{n}_1\equiv\frac{N}{A}$, have been
introduced. The expression of the temperature given by
Eq. (\ref{Tbath}) differs from the expression given in \cite{mgb19} by
a factor $4$ because here the two walls are vibrating. 

\begin{figure}
\begin{center}
\includegraphics[angle=0,width=0.65\linewidth,clip]{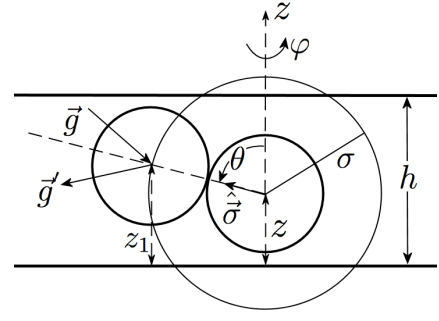}
\end{center}
%\vspace*{-2cm}
\caption{Collision between the tagged particle and a bath particle. 
$\theta$ and $\varphi$ are the polar and
azimuthal angles respectively. }\label{confinadoFig}
\end{figure}

The objective now is to study the dynamics of the tagged particle. It 
will be assumed that the collisions between the tagged particle and the
ones of the bath do not modify the state of the bath. The
evolution equation for the one-particle distribution function of the
tagged particle, $f$, 
immersed in the bath described by the one-particle distribution
function, $f_1$, follows by the same arguments used to derive the
Boltzmann equation for confined systems \cite{bmg16, bgm17, mgb19, bgm20} and
the following Boltzmann-Lorentz like equation is obtained 
\begin{eqnarray}\label{bleqz}
\left(\frac{\partial}{\partial
  t}+\mathbf{v}\cdot\frac{\partial}{\partial\mathbf{r}}\right)
f(\mathbf{r},\mathbf{v},t)
=J_z[f_1|f]
+L_W f(\mathbf{r},\mathbf{v},t). 
\end{eqnarray}
Here $J_z$ is the collisional contribution that takes into account the
collisions between the tagged particle and the particles of the bath, 
\begin{eqnarray}
J_z[f_1|f]=\sigma^2\int
d\mathbf{v}_1\int_{\Sigma(z)}d\sig\abs{\mathbf{g}\cdot\sig}
[\Theta(\mathbf{g}\cdot\sig)\alpha^{-2}b_{\sig}^{-1}-\Theta(-\mathbf{g}\cdot\sig)]\nonumber\\
f_1(\mathbf{v}_1)f(\mathbf{r}, \mathbf{v},t), \nonumber\\
\end{eqnarray}
where we have introduced the Heaviside step function, $\Theta$, the
operator $b_{\sig}^{-1}$ that replaces all velocities appearing to its
right by the precollisional velocities $\mathbf{v}^*$ and
$\mathbf{v}_1^*$, 
\begin{eqnarray}
\mathbf{v}^*\equiv
  b_{\sig}^{-1}\mathbf{v}=\mathbf{v}+\frac{m}{m+M}(1+\alpha^{-1})
(\mathbf{g}\cdot\sig)\sig, \\
\mathbf{v}_1^*\equiv
  b_{\sig}^{-1}\mathbf{v}_1=\mathbf{v}_1-\frac{M}{m+M}(1+\alpha^{-1})
(\mathbf{g}\cdot\sig)\sig,  
\end{eqnarray}
and the region of integration of $\sig$, $\Sigma(z)$, which depends on
the confinement. In spherical coordinates, $d\sig=\sin\theta
d\theta d\varphi$, where $\theta$ and $\varphi$ are the polar and
azimuthal angles respectively (see Fig. \ref{confinadoFig}) and the set $\Sigma$ can
be parametrized as
\begin{equation}
\Sigma(z)=\left\{(\theta, \varphi)|\theta\in\left(\frac{\pi}{2}-b_2(z), \frac{\pi}{2}+b_1(z)
    \right), \varphi\in(0,2\pi)\right\}, 
\end{equation}
with
\begin{eqnarray}
b_1(z)=\arcsin\left(\frac{z-\sigma/2}{\sigma}\right), \\
b_2(z)=\arcsin\left(\frac{h-z-\sigma/2}{\sigma}\right). 
\end{eqnarray}
Finally, the wall contribution is \cite{dorfVBeij}
\begin{eqnarray}
L_Wf(\mathbf{r},\mathbf{v},t)=[\delta(z-\sigma/2)L_b+\delta(z-h+\sigma/2)L_t]
f(\mathbf{r},\mathbf{v},t), \nonumber\\
\end{eqnarray}
with
\begin{eqnarray}
L_bf(\mathbf{r},\mathbf{v},t)=[\Theta(v_z-2v_0)\abs{2v_0-v_z}b_b-\Theta(-v_z)\abs{v_z}]
f(\mathbf{r},\mathbf{v},t), \nonumber\\\\
L_tf(\mathbf{r},\mathbf{v},t)=[\Theta(-v_z-2v_0)\abs{2v_0+v_z}b_t-\Theta(v_z)v_z]
f(\mathbf{r},\mathbf{v},t). \nonumber\\
\end{eqnarray}
In contrast with the ``traditional'' Boltzmann-Lorentz equation, the
integration in $\sig$ is restricted to $\Sigma(z)$ because, otherwise,
the particle of the bath that collides with the tagged particle would
not fulfill the constraint of being confined between the two walls. 

As in the case of the bath, it is a good approximation to neglect the
$z$ dependence of $f$. Then
\begin{equation}
f(\mathbf{r},\mathbf{v},t)\approx
f(\mathbf{r}_{\perp},\mathbf{v},t)\equiv\frac{1}{h-\sigma}
\int_{\sigma/2}^{h-\sigma/2}dzf(\mathbf{r},\mathbf{v},t), 
\end{equation}
where we have introduced the perpendicular component to the
$z$-direction of a vector through 
$\mathbf{a}_{\perp}\equiv a_x\mathbf{e}_x+a_y\mathbf{e}_y$. In this
situation, by integrating over $z$ in Eq. (\ref{bleqz}) and replacing 
$f(\mathbf{r},\mathbf{v},t)$ by $f(\mathbf{r}_{\perp},\mathbf{v},t)$ in
the collisional operator, $J_z$, it is obtained
\begin{eqnarray}\label{bleq}
&&\left(\frac{\partial}{\partial
  t}+\mathbf{v}_{\perp}\cdot\frac{\partial}{\partial\mathbf{r}_{\perp}}\right)
f(\mathbf{r}_{\perp},\mathbf{v},t) \nonumber\\
&&=\frac{1}{h-\sigma}\int_{\sigma/2}^{h-\sigma/2}dzJ_z[f_1|f]
+\frac{1}{h-\sigma} (L_b+L_t) f(\mathbf{r}_{\perp},\mathbf{v},t),  \nonumber\\
\end{eqnarray}
that is a closed evolution equation for
$f(\mathbf{r}_{\perp},\mathbf{v},t)$. Note that, although the $z$
variable does not appear in Eq. (\ref{bleq}), the component $v_z$ still
remains.

\section{Dynamics of spatially homogeneous states}\label{homoStates}
In this section, we will focus on the study of the dynamics of the
tagged particle in the most simple situation, in which the system can
also be considered spatially homogeneous. In this case, the
one-particle distribution function does not depend on
$\mathbf{r}_{\perp}$ and Eq. (\ref{bleq}) leads to 
\begin{eqnarray}\label{bleqApproxHom}
\frac{\partial}{\partial
  t}f(\mathbf{v},t)
=\frac{1}{h-\sigma}\int_{\sigma/2}^{h-\sigma/2}dzJ_z[f_1|f]
+\frac{1}{h-\sigma} (L_b+L_t) f(\mathbf{v},t). \nonumber\\
\end{eqnarray}
This equation is still difficult to deal with, and we will
further assume that $f$ can be approximated by a two-temperatures gaussian
distribution, i.e.
\begin{equation}\label{dfGaussian}
f(\mathbf{v},t)=\frac{n}{\pi^{3/2}\Omega(t)^2\Omega_z(t)^2}
\exp\left[-\frac{v_x^2+v_y^2}{\Omega(t)^2}-\frac{v_z^2}{\Omega_z(t)^2}\right], 
\end{equation}
with $n\equiv\frac{1}{(h-\sigma)A}$. The thermal velocities, $\Omega(t)$
and $\Omega_z(t)$, are related to the 
horizontal, $T(t)$, and vertical, $T_z(t)$, temperatures through
\begin{eqnarray}
\frac{M}{2}\Omega^2(t)=T(t), \quad \frac{M}{2}\Omega_z(t)^2=T_z(t),  
\end{eqnarray}
where the horizontal and vertical temperatures are defined as in
Eq. (\ref{defTTempsBa}) for the bath 
\begin{equation}\label{defTemps}
nT(t)=\frac{M}{2}\int d\mathbf{v}(v_x^2+v_y^2)
f(\mathbf{v},t), \quad
nT_z(t)=M\int
d\mathbf{v}v_z^2f(\mathbf{v},t). 
\end{equation}
The validity of the simple ansatz given by
Eq. (\ref{dfGaussian}) will be confirmed by Molecular Dynamics (MD)
simulation results, at least for some range of the system
parameters. 

Closed evolution equations for the horizontal and vertical
temperatures are obtained by taking velocity moments in
Eq. (\ref{bleqApproxHom}). For simplicity, we will write the
equivalent evolution equations for $\Omega^2$ and $\Omega_z^2$. By
multiplying Eq. (\ref{bleqApproxHom}) by 
$(v_x^2+v_y^2)$ and by $v_z^2$ followed by integrating in the velocity
space, it is obtained 
\begin{eqnarray}\label{ecEvT}
\frac{d}{dt}\Omega^2&=&\mathcal{G}(\Omega^2, \Omega_z^2), \\
\frac{d}{dt}\Omega_z^2&=&\mathcal{H}(\Omega^2, \Omega_z^2)
+\frac{4v_0}{\epsilon\sigma}\Omega_z^2. \label{ecEvTz}
\end{eqnarray} 
The collisional terms are given by 
\begin{eqnarray}
\mathcal{G}(\Omega^2, \Omega_z^2)&=&\frac{1}{(h-\sigma)n}\int
d\mathbf{v}(v_x^2+v_y^2)\int_{\sigma/2}^{h-\sigma/2}dzJ_z[f_1|f], \label{mathGeq}
\nonumber\\
\\
\mathcal{H}(\Omega^2, \Omega_z^2)&=&\frac{2}{(h-\sigma)n}\int
d\mathbf{v}v_z^2\int_{\sigma/2}^{h-\sigma/2}dzJ_z[f_1|f], \label{mathHeq}
\end{eqnarray}
and it has been used the exact result derived in \cite{br09} that
establishes that the energy injected by the walls is $v_0$ times the
pressure. Note that, since energy is injected in the
vertical direction, collisions with the walls only contribute to
the vertical thermal velocity equation. The collisional terms $\mathcal{G}$ and
$\mathcal{H}$ are evaluated in 
Appendix \ref{velMomen}, obtaining
%\begin{eqnarray}
%\mathcal{G}=\sqrt{\pi}(1+\alpha)
%n_1n\sigma^2 \frac{mM}{\epsilon(m+M)}\nonumber\\\left\{ (1+\alpha) \frac{m}{m+M}
%\int_0^{\epsilon}
%  dy(\epsilon-y)(1-y^2)[(w^2+\Omega^2)(1-y^2)+(w_z^2+\Omega_z^2)y^2]^{3/2}\right.
%\nonumber\\
%\left. -2\Omega^2\int_0^{\epsilon}dy(\epsilon-y)(1-y^2)
%[(w^2+\Omega^2)(1-y^2)+(w_z^2+\Omega_z^2)y^2]^{1/2}\right\}, 
%\end{eqnarray}
\begin{eqnarray}\label{ecG}
&&\mathcal{G}=2\sqrt{\pi}
n_1\sigma^2\frac{\mu}{\epsilon}\int_0^{\epsilon}
  dy(\epsilon-y)(1-y^2)\nonumber\\
&&\left\{
\mu
[(w^2+\Omega^2)(1-y^2)+(w_z^2+\Omega_z^2)y^2]^{3/2}\right.\nonumber\\
&&\left. -2\Omega^2
[(w^2+\Omega^2)(1-y^2)+(w_z^2+\Omega_z^2)y^2]^{1/2}\right\}, 
\end{eqnarray}
and 
\begin{eqnarray}\label{ecH}
&&\mathcal{H}=4\sqrt{\pi}
n_1\sigma^2 \frac{\mu}{\epsilon}\int_0^{\epsilon}
  dy(\epsilon-y)y^2\nonumber\\
&&\left\{
\mu
[(w^2+\Omega^2)(1-y^2)+(w_z^2+\Omega_z^2)y^2]^{3/2}\right.\nonumber\\
&&\left. -2\Omega_z^2
[(w^2+\Omega^2)(1-y^2)+(w_z^2+\Omega_z^2)y^2]^{1/2}\right\}, 
\end{eqnarray}
where the dimensionless parameter
\begin{equation}
\mu=\frac{m}{m+M}(1+\alpha), 
\end{equation}
has been introduced. The
integrals given by Eqs. (\ref{ecG}) and (\ref{ecH}) can be 
evaluated exactly, but their expressions are very long and we prefer to
leave them in the more compact form given above. Nevertheless, some
relatively simpler expressions are obtained for thin
systems, by expanding the expressions of $\mathcal{G}$ and
$\mathcal{H}$ to second order in $\epsilon$, 
\begin{eqnarray}\label{ecGApp}
\mathcal{G}\approx 2\mu\,\nu%\nonumber\\
\left\{\frac{\mu}{2}\left[\left(1-\frac{5}{12}\epsilon^2 \right)(w^2+\Omega^2)
+\frac{\epsilon^2}{4} (w_z^2+\Omega_z^2)\right]\right.\nonumber\\
\left.-\frac{\Omega^2}{w^2+\Omega^2}
\left[\left(1-\frac{\epsilon^2}{4}\right)(w^2+\Omega^2)
+\frac{\epsilon^2}{12} (w_z^2+\Omega_z^2)\right]
\right\}, \nonumber\\
\end{eqnarray}
and 
\begin{eqnarray}\label{ecHApp}
\mathcal{H}\approx\frac{\epsilon^2}{3}
\mu\,\nu
\left[\mu(w^2+\Omega^2)-2\Omega_z^2\right]. 
\end{eqnarray}
In the above expressions  
\begin{equation}
\nu=\sqrt{\pi}\tilde{n}_1\sigma(w^2+\Omega^2)^{1/2}, 
\end{equation}
that, to leading order in $\epsilon$, is proportional to the collision frequency of the tagged
particle. 

Eqs. (\ref{ecEvT}) and (\ref{ecEvTz}) with $\mathcal{G}$ and
$\mathcal{H}$ given by Eqs. (\ref{ecG}) and (\ref{ecH}) respectively
(or their approximate expressions to $\epsilon^2$ given in
Eqs. (\ref{ecGApp}) and (\ref{ecHApp})) form a closed system of
differential equations for the horizontal and vertical
thermal velocities. Let us stress that all the dependence in the
masses, $m$ and $M$, and in the 
inelasticity of the tagged particle, $\alpha$, in the evolution
equations goes through the parameter $\mu$. This is similar to what
happens in the non-confined free cooling case \cite{bds99,sd01,sd01b}. The system
of differential equations is highly non-linear, but its structure is
clear: the collisions with the walls inject energy in the vertical
direction, while the collisions with the bath particles
inject/dissipate energy in the vertical and horizontal directions. The
collisional contribution to the horizontal thermal velocity is given by
$\mathcal{G}$ and, to leading order in $\epsilon$, it is 
\begin{equation}\label{ecGApp2}
\mathcal{G}\approx\mathcal{G}_0=\mu\,\nu
\left[\mu(w^2+\Omega^2)-2\Omega^2\right],  
\end{equation}
that is the same as that for inelastic collisions in two dimensions
\cite{sd01,sd01b}. To leading order in $\epsilon$, the collisional contribution to the
vertical thermal velocity is given by Eq. (\ref{ecHApp}). Its structure
is similar to that of Eq. (\ref{ecGApp2}), but multiplied by the
geometrical factor $\epsilon^2/3$. This can be intuitively understood
as, the thinner the system, the slowest the dynamics of $\Omega_z$
is. In addition, $\Omega$ is replaced by $\Omega_z$ 
in the ``friction'' term that leads to equipartition in the elastic
case with the elastic walls at rest, i.e. $v_0=0$. 

Before embarking in the analysis of Eqs. (\ref{ecEvT}) and
(\ref{ecEvTz}), let us consider a simpler situation which leads to a
system of differential equations that can be analytically solved and
that will help us to understand many (if not all) features of the
general case. If $\Omega_z/\Omega$ is not very large, $\mathcal{G}$
can be approximated by its leading order in $\epsilon$ 
contribution, i.e. $\mathcal{G}\approx\mathcal{G}_0$. 
This simplifies considerably the analysis, as the dynamics of $\Omega$ is
decoupled from $\Omega_z$ within this approximation. It is convenient
to introduce the dimensionless thermal velocities
\begin{equation}
X\equiv\frac{\Omega^2}{w^2}, \quad Y\equiv\frac{\Omega_z^2}{w^2}, 
\end{equation}
and the dimensionless time, $\tau$, through
\begin{equation}
d\tau=\nu dt, 
\end{equation}
that, to leading order in $\epsilon$, is proportional to the number of
collisions the tagged particle experiments in the time interval
$(0,t)$. In this time scale, the evolution equations are 
\begin{eqnarray}
\frac{d}{d\tau}X&=&-\mu(2-\mu)X+\mu^2\label{eqOApp}\\
\frac{d}{d\tau}Y&=&
-\frac{2}{3}\left[\mu-\frac{1}{K\sqrt{2(1+X)}}\right]
\epsilon^2Y+\frac{\mu^2}{3}\epsilon^2(1+X), \nonumber\\\label{eqOzApp}
\end{eqnarray}
where 
\begin{equation}\label{DefK}
K(\alpha_1,
\epsilon)=\frac{\gamma_1}{(1+\alpha_1)\left(\gamma_1-\frac{1+\alpha_1}{2}
\right)},  
\end{equation}
is a function depending on the inelasticity of the particles of the
bath, $\alpha_1$, and on $\epsilon$ (it does not depend on the
inelasticity of the tagged particle, $\alpha$). Note that, in these
units, the dynamics is independent of the walls velocity and all the
dependence on the inelasticity of the bath particles comes through
$K$. Eq. (\ref{eqOApp}) is 
an inhomogeneous linear equation for $X$ and the time scale in
which it evolves is of the order of $\mu^{-1}$. On the other
hand, in Eq. (\ref{eqOzApp}) $Y$ is coupled with $X$, but the
time scale in which $Y$ evolves is of the order of
$(\mu\epsilon^2)^{-1}$, so that, in this time scale, it can be assumed
that $X$ instantaneously reaches its stationary value,
$X_s$, given by 
\begin{equation}\label{omegaSt}
X_s=\frac{\mu}{2-\mu}. 
\end{equation}
By substituting $X$ by $X_s$ given by
Eq. (\ref{omegaSt}) in Eq. (\ref{eqOzApp}), 
the following approximate equation for $Y$ is obtained 
\begin{equation}
\frac{d}{d\tau}Y=
-\frac{2}{3}\left[\mu-\frac{(2-\mu)^{1/2}}{2K}\right]
\epsilon^2Y+\frac{2\mu^2}{3(2-\mu)}\epsilon^2. 
\end{equation}
If $\mu-\frac{(2-\mu)^{1/2}}{2K}>0$, $Y$ reaches the
following stationary value
\begin{equation}\label{omegaZSt}
Y_s=\frac{2\mu^2K}{(2-\mu)[2\mu K-(2-\mu)^{1/2}]}. 
\end{equation} 
Otherwise, $Y$ diverges and there is not a stationary
state. Hence, a critical value of $\mu$, $\mu_c$, can be identified as 
\begin{equation}\label{muCApp}
\mu_c=\frac{\sqrt{1+32K^2}-1}{8K^2}. 
\end{equation}
For $\mu<\mu_c$, there is not a stationary state. Note that $\mu_c$ depends
on $\epsilon$ and the inelasticity of the particles of the bath,
$\alpha_1$, but it is independent of the inelasticity of the tagged
particle. In fact, $\lim_{\mu\to\mu_c^+}Y_s=\infty$, $X_s$
remaining finite.  Equivalently, the
critical value of the mass, $M_c$, above which there is not a
stationary state is
\begin{equation}\label{McAppEq}
\frac{M_c}{m}=\frac{(1+\alpha)8K^2}{\sqrt{1+32K^2}-1}-1. 
\end{equation}
To summarize, the dynamics of the tagged particle in these conditions 
consists of a fast equilibration in the horizontal direction followed
by a slow evolution of $\Omega_z$, that eventually will reach its
stationary value if $\mu>\mu_c$. This can be
intuitively understood as, for the considered geometry, horizontal
collisions (the ones that stabilize $\Omega$) are much more probable
than collisions in the vertical 
direction. The origin of the instability can also be understood. In
effect, from Eq. (\ref{ecEvTz}) it is seen that the wall contribution
is $\mu$-independent, while the collisional contribution increase with
$\mu$ (consistently with the fact that for more massive tagged
particle, less efficient the collisional contribution is). So, for
small enough $\mu$ the ``friction'' mechanism is not able to
compensate the energy injection and $\Omega_z$ diverges. 

Let us consider now the general case given by Eqs. (\ref{ecEvT}) and
(\ref{ecEvTz}) with $\mathcal{G}$ and $\mathcal{H}$ given by their
second order in $\epsilon$ expressions (Eqs. (\ref{ecGApp}) and
(\ref{ecHApp})). In this case, if $\Omega_z^2$ is much larger than
$\Omega^2$, as it is the case close to the critical mass, $\Omega^2$
is no longer a fast variable due to the coupling $\epsilon^2
\Omega_z^2$. This coupling may affect the values of the stationary
values as long as the critical value of the tagged mass. In effect,
from Eqs. (\ref{ecEvT}) and (\ref{ecEvTz}), 
the stationary values, $\Omega_s^2$ and $\Omega_{z,s}^2$, fulfill
the following set of two equations 
\begin{eqnarray}
\mathcal{G}(\Omega_s^2, \Omega_{z,s}^2)=0, \label{EcGst}\\
\mathcal{H}(\Omega_s^2,
  \Omega_{z,s}^2)+\frac{4v_0}{\epsilon\sigma}\Omega_z^2=0.   
\label{EcHst}
\end{eqnarray}
Although we are not going to write it explicitly, the dimensionless
thermal velocities, $X$ and $Y$, in the $\tau$ scale, also verify a
system of differential equations in which $v_0$ can be scaled. In
fact, the system of Eqs. (\ref{EcGst}) and (\ref{EcHst}) for the
stationary thermal velocities can be transformed in the
following one for $X_s$ and $Y_s$
\begin{eqnarray}
Y_s
&=&\frac{\mu^2(1+X_s)}
{2\mu-\frac{\sqrt{2}}{K\sqrt{1+X_s}}}, 
\label{ecOmS}\\
Y_s&=&\frac{\left[\left(1-\frac{\epsilon^2}{4}\right)X_s
-\frac{\mu}{2}\left(1-\frac{5\epsilon^2}{12}\right)(1+X_s)\right]
(1+X_s)}
{\frac{\epsilon^2}{4}\left[\frac{\mu}{2}(1+X_s)-\frac{X_s}{3}\right]}
%(w^2+\Omega_s^2)
-\gamma_1, \nonumber\\\label{ecOmZS}
\end{eqnarray}
that leads to a quintic equation that
can be solved numerically. For the considered values of the
parameters, as in the the approximate
analysis made before, there is only one 
physical solution if $\mu>\mu_c$. Moreover,
$\lim_{\mu\to \mu_c^+}Y_s=\infty$, $X_s$ remaining
finite. If $\mu<\mu_c$, there is no physical solution. The explicit
expression of $\mu_c$ for Eqs. (\ref{ecOmS}) and  (\ref{ecOmZS})
is given by imposing the divergence of $Y_s$, i.e. 
\begin{eqnarray}
\mu_c-\frac{1}{K\sqrt{2(1+X_s)}}&=&0, \\
\frac{\mu_c}{2}(1+X_s)-\frac{X_s}{3}&=&0, \label{omegaScritico}
\end{eqnarray}
whose  solution is
\begin{equation}\label{muC}
\mu_c=\frac{\sqrt{9+32K^2}-3}{8K^2}, 
\end{equation}
where $K$ is given by Eq. (\ref{DefK}). 
This expression differs from the approximation obtained
previously, Eq. (\ref{muCApp}), but they agree when the inelasticity
of the particles of the bath tends to the elastic limit because 
$\lim_{\alpha_1\to 1}K(\alpha_1,\epsilon)=\infty$ and
$\mu_c\approx\frac{1}{\sqrt{2}K}$ in both cases. The explicit
  expression of the critical mass in the context of
  Eqs. (\ref{ecOmS}) and  (\ref{ecOmZS}) is 
\begin{equation}\label{McEq}
\frac{M_c}{m}=\frac{8(1+\alpha)K^2}
{\sqrt{9+32K^2}-3}-1.  
\end{equation}

\begin{figure}
\begin{center}
\includegraphics[angle=0,width=0.85\linewidth,clip]{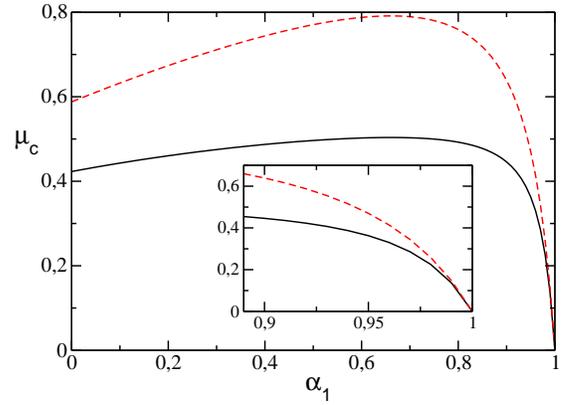}
\end{center}
%\vspace*{-2cm}
\caption{(Color online) $\mu_c$ as a function of the
inelasticity of the bath particles, $\alpha_1$, for
$\epsilon=0.5$. The solid line is the theoretical prediction given by
Eq. (\ref{muC}) and the (red) dashed line is the approximate
expression given by Eq. (\ref{muCApp}). In the inset, the region
close to the elastic limit is shown. }\label{mucFig}
\end{figure}
\begin{figure}
\begin{center}
\includegraphics[angle=0,width=0.85\linewidth,clip]{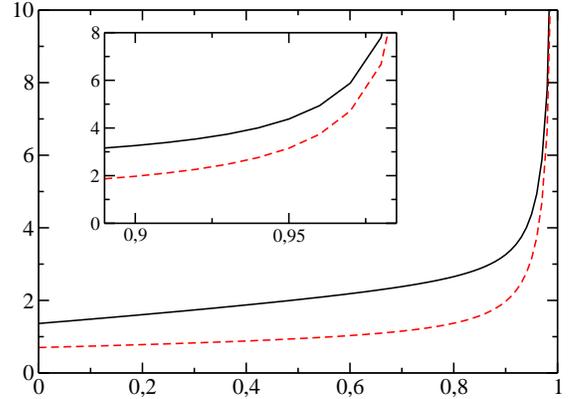}
\end{center}
%\vspace*{-2cm}
\caption{(Color online) Critical mass as function of the inelasticity for
  $\epsilon=0.5$. It has been considered that $\alpha=\alpha_1$. The
  solid line is the theoretical prediction given 
by Eq. (\ref{McEq}) and the (red) dashed line the approximate
expression given by Eq. (\ref{McAppEq}). In the inset, the region
close to the elastic limit is shown. }\label{mc05Fig}
\end{figure}

In Fig. \ref{mucFig}, $\mu_c$ is plotted as a function of the
inelasticity of the bath particles, $\alpha_1$, for
$\epsilon=0.5$. The solid line is the theoretical prediction given by
Eq. (\ref{muC}) and the (red) dashed line is the approximate
expression given by Eq. (\ref{muCApp}). It is seen that the
approximate value is always larger than the exact (up to $\epsilon^2$
order) value and that both agree in the elastic limit. This can be
understood from the fact that, at the critical 
point, $X_s$ is given by Eq. (\ref{omegaScritico}), i.e. 
$X_{s,c}=\frac{\mu}{\frac{2}{3}-\mu}$, which is larger than
the one given by Eq. (\ref{omegaSt}) and which renormalize
$\mu_c$ into a smaller value. Similar results are obtained for other
values of the separation between the walls. To have a clearer physical picture, in
Fig. \ref{mc05Fig}, we have 
plotted the critical mass for $\epsilon=0.5$. As this quantity also depends on
the inelasticity of the tagged particle, we have considered the case
$\alpha=\alpha_1$. The solid line is the theoretical prediction given
by Eq. (\ref{McEq}) and the (red) dashed line the approximate
expression given by Eq. (\ref{McAppEq}). It is seen that the critical mass diverges in the
elastic limit but, remarkably, for mild inelasticities, let us say
till $\alpha\approx 0.9$, the critical mass is smaller than $3m$, so
that the instability is developed ``very soon''. 

\section{Simulation results}\label{secSR}

In this section we present MD simulation results of the model
introduced in Sec. \ref{secModel} in order to compare
them with the theoretical predictions obtained in the previous
section. The MD simulations are performed using the event-driven
algorithm \cite{allen} taking $m$, $\sigma$ and $v_0$ as units of
mass, length and velocity respectively. 
%{\bf The used parameters are $N=585$,
%$\tilde{n}_1\sigma^2=0.06$, $\alpha=\alpha_1=0.98$ and $\epsilon=0.5$,
%varying the
%tagged particle mass. Note that, for the above mentioned parameters, 
%$\frac{M_c}{m}\approx 7.8$.} 
The used parameters for all the simulations are $N=585$,
$\tilde{n}_1\sigma^2=0.06$, and $\epsilon=0.5$,
varying the tagged particle mass and the coefficients of normal restitution,
$\alpha$ and $\alpha_1$. The initial condition is generated by putting the
particles of the bath with a two-temperatures maxwellian corresponding
to the theoretical prediction and the tagged particle at rest. In most
of the simulations, the results have been averaged over $20$ trajectories. If not, it is
explicitly indicated. We have also seen that the bath is always
spatially homogeneous and we have controlled if it was disturbed by
the presence of the intruder. 

\begin{figure}[h]
\begin{center}
\includegraphics[angle=0,width=0.85\linewidth,clip]{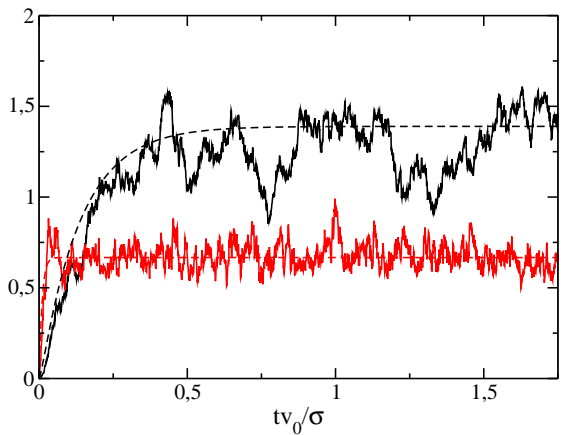}
\end{center}
%\vspace*{-2cm}
\caption{(Color online) $X$ and $Y$ as a function of the dimensionless time,
  $v_0t/\sigma$, for $M=1.5m$. The (black and red)
    solid lines are the simulation results for $Y$and $X$ 
 (respectively) and the (black and red) dashed lines are the (corresponding)
  theoretical predictions. }
\label{xyEvolucionFig}
\end{figure}
\begin{figure}[h]
\begin{center}
\includegraphics[angle=0,width=0.85\linewidth,clip]{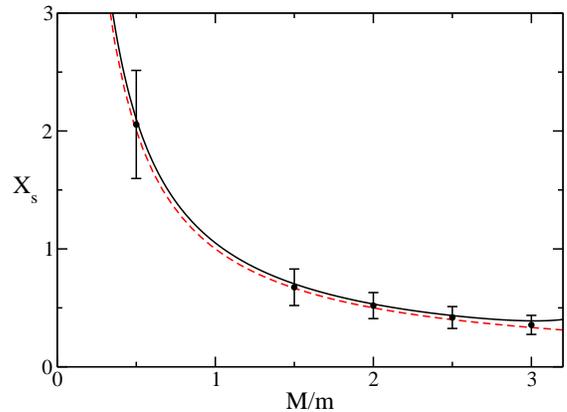}
\end{center}
%\vspace*{-2cm}
\caption{(Color online) Stationary values of  the dimensionless horizontal thermal
  velocity, $X_s$, as a
  function of the dimensionless mass of the intruder, $M/m$. The
  points are the simulation results and the solid line the theoretical
  prediction given by the numerical 
solution of the system of equations (\ref{ecOmS}) and
(\ref{ecOmZS}). The (red) dashed line is the approximate solution given by
Eq. (\ref{omegaSt}).}
\label{xsFig}
\end{figure}
\begin{figure}[h]
\begin{center}
\includegraphics[angle=0,width=0.85\linewidth,clip]{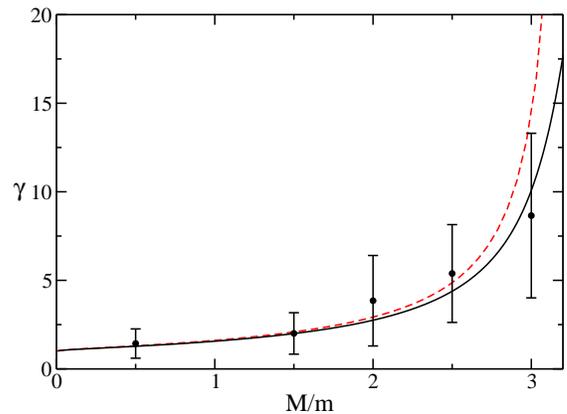}
\end{center}
%\vspace*{-2cm}
\caption{(Color online) Quotient between the
stationary temperatures, 
$\gamma\equiv\frac{Y_s}{X_s}$, as a function of
$\frac{M}{m}$. The dots are the simulation results and the solid line
the theoretical prediction given by the numerical solution of the
system of equations (\ref{ecOmS}) and (\ref{ecOmZS}). The (red) dashed line
is the approximate solution given by Eqs. (\ref{omegaSt}) and
(\ref{omegaZSt}) }
\label{ysFig}
\end{figure}

In first place, we have considered a system with $\alpha_1=0.95$ and
  $\alpha=1.0$. For these values of the parameters
  $\frac{M_c}{m}\approx 4.5$.  In Fig. \ref{xyEvolucionFig} the
  dimensionless horizontal and vertical 
temperatures, $X$ and $Y$, are plotted as a function of the
dimensionless time, $v_0t/\sigma$, for
$\frac{M}{m}=1.5$. The solid lines are
  the simulation results averaged over $100$
  realizations for $Y$and $X$. $Y$ reaches a larger stationary value, as
  expected. The dashed lines are the numerical solution of
Eqs. (\ref{ecEvT}) and (\ref{ecEvTz}) with 
  $\mathcal{G}$ and $\mathcal{H}$ given by their expression to second
  order in $\epsilon$, Eqs. (\ref{ecGApp}) and (\ref{ecHApp}). In this
  case, if $\mathcal{G}$ is further approximated by $\mathcal{G}_0$,
  an indistinguishable result is obtained. It can be observed that, as
  discussed in Sec. \ref{homoStates}, the horizontal temperature reaches the
  stationary value much quicker than the vertical temperature and that the
  agreement between the theoretical prediction and the simulation
  results is remarkably good. In this case, we have also checked that
  the bath parameters are not disturbed by the presence of the
  intruder, finding that the distribution function of the bath is
  approximately a two-temperatures gaussian with the measured
  temperatures in agreement with the theoretical predictions given by
  Eqs. (\ref{gamma1}) and (\ref{Tbath}). 

Similar results can be obtained for different values of the tagged
particle mass, from which the stationary values of the horizontal and
vertical temperatures, as long as their corresponding error bars can be
easily measured. In fact, also with  
$\frac{M}{m}\approx 3.5$, we have seen that the parameters of the bath
are not disturbed by the presence of the intruder. For $\frac{M}{m}>3.5$, 
the bath velocity distribution function starts deviating from the gaussian
and the partial temperatures from their theoretical predictions. 
In Fig. \ref{xsFig}, the dimensionless stationary horizontal
temperature is plotted as a function of the dimensionless mass of the
tagged particle, $\frac{M}{m}$. The dots are the simulation results
and the solid line the theoretical prediction given by the numerical
solution of the system of equations (\ref{ecOmS}) and
(\ref{ecOmZS}). The (red) dashed line is the approximate solution given by
Eq. (\ref{omegaSt}). It can be seen that the agreement between the
simulation results and the theoretical prediction is good, being
the two theoretical predictions very similar for   
$\frac{M}{m}\lesssim 3$. In Fig. \ref{ysFig}, the quotient between the
stationary temperatures, 
$\gamma\equiv\frac{Y_s}{X_s}$, is plotted as a function of
$\frac{M}{m}$. The dots are the simulation results and the solid line
the theoretical prediction given by the numerical solution of the
system of equations (\ref{ecOmS}) and (\ref{ecOmZS}). The (red) dashed line
is the approximate solution given by Eqs. (\ref{omegaSt}) and
(\ref{omegaZSt}). In this case, for $\frac{M}{m}\approx 3$, there are
already some differences between the two theoretical predictions,
being the simulation results close to the former, as expected. Again,
the agreement between the simulation results and the theoretical
prediction is very good. 

\begin{figure}
\begin{center}
\includegraphics[angle=0,width=0.85\linewidth,clip]{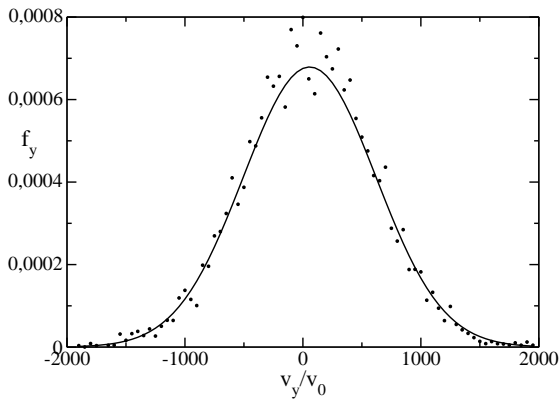}
\end{center}
%\vspace*{-2cm}
\caption{ Normalized marginal velocity
distribution in the $y$ direction, $f_y$, as a function of
the dimensionless velocity, $\frac{v_y}{v_0}$, for $M=4m$. The points are the
simulation results and the solid line the gaussian approximation.  }\label{fyvyM4Fig}
\end{figure}

\begin{figure}
\begin{center}
\includegraphics[angle=0,width=0.85\linewidth,clip]{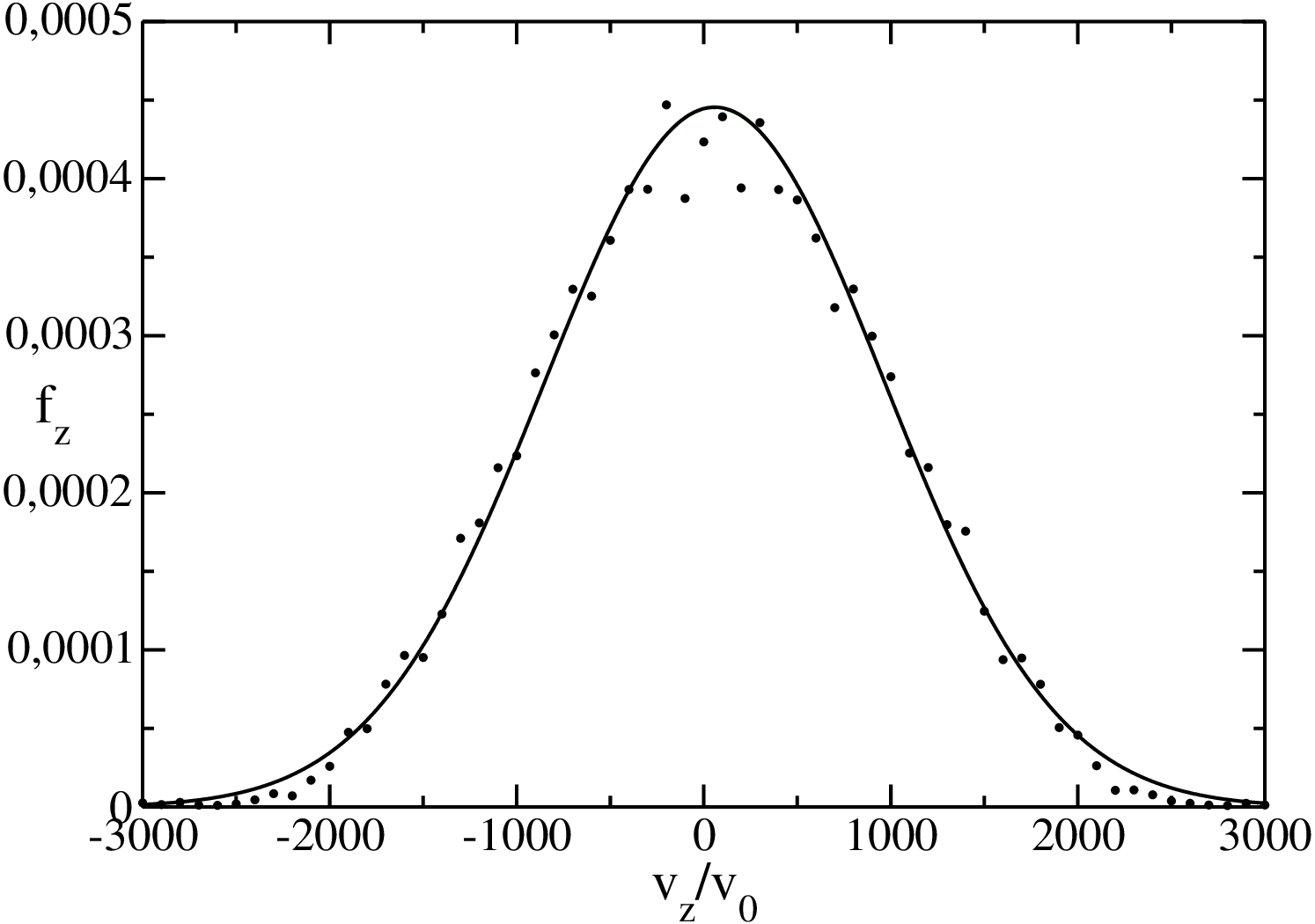}
\end{center}
%\vspace*{-2cm}
\caption{Normalized marginal velocity
distribution in the $z$ direction, $f_z$, as a function of
the dimensionless velocity, $\frac{v_z}{v_0}$, for $M=4m$. The points are the
simulation results and the solid line the gaussian approximation }\label{fzvzM4Fig}
\end{figure}

\begin{figure}
\begin{center}
\includegraphics[angle=0,width=0.85\linewidth,clip]{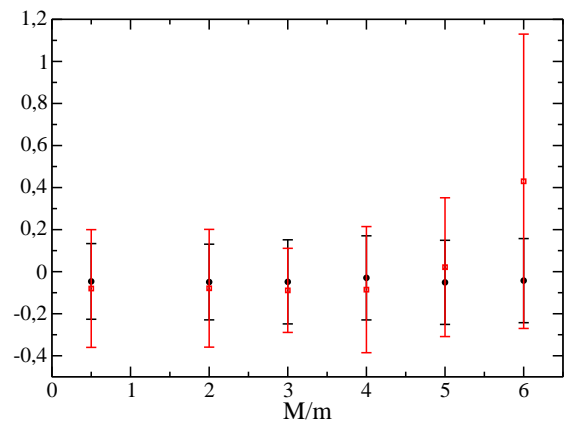}
\end{center}
%\vspace*{-2cm}
\caption{(Color online) $a_{2,xy}$ and $a_{2,z}$ as a function of the
  dimensionless mass, $\frac{M}{m}$. The (black) circles are the
  simulation results for $a_{2,xy}$ and the (red) squares for
  $a_{2,z}$. }\label{kurtosisFig} 
\end{figure}

In the following, we present simulations
results for a system with $\alpha=\alpha_1=0.98$ where
$\frac{M_c}{m}\approx 7.8$. We have performed the same analysis as
before, finding similar results \cite{gmb22} and we have controlled
that the intruder velocity distribution function is approximately gaussian.
In Fig. \ref{fyvyM4Fig} the
normalized marginal velocity 
distribution in the $y$ direction, $f_y$, is plotted as a function of
the dimensionless velocity, $\frac{v_y}{v_0}$, for $M=4m$. The points are the
simulation results and the solid line the gaussian approximation. The
same is plotted in Fig. \ref{fzvzM4Fig} but in the $z$ direction. It
is observed that the gaussian approximation accurately describes the
shape of the marginal distributions, at least for thermal velocities
where the data are shown. Similar results are obtain for $M<5m$. 
A more quantitative analysis can be carried out by 
measuring the kurtosis of the
marginal distributions
\begin{equation}
a_{2,xy}=\frac{\langle(v_x^2+v_y^2)^2\rangle}{2\langle
  v_x^2+v_y^2\rangle^2}-1, \quad a_{2,z}=\frac{\langle
  v_z^4\rangle}{3\langle v_z^2\rangle^2}-1, 
\end{equation}
where $\langle\dots\rangle$ means average over different realizations
in the stationary state. In Fig. \ref{kurtosisFig}, the (black)
circles and the (red) squares are the simulation results for $a_{2,xy}$ and
$a_{2,z}$ respectively, that are plotted as a function of the
dimensionless mass. It is observed that $a_{2,xy}$ remains
approximately unchanged for the plotted mass values, while $a_{2,z}$
start varying with respect to the small-mass value at $M\approx
5m$, strongly deviating from the gaussian value already for $M=6$. 

\begin{figure}
\begin{center}
\includegraphics[angle=0,width=0.85\linewidth,clip]{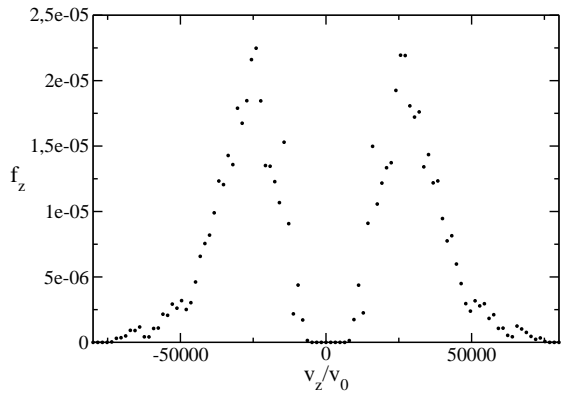}
\end{center}
%\vspace*{-2cm}
\caption{Normalized marginal velocity
distribution in the $z$ direction, $f_z$, as a function of
the dimensionless velocity, $\frac{v_z}{v_0}$, for $M=10m$. }\label{fzvzM10Fig}
\end{figure}

\begin{figure}
\begin{center}
\includegraphics[angle=0,width=0.98\linewidth,clip]{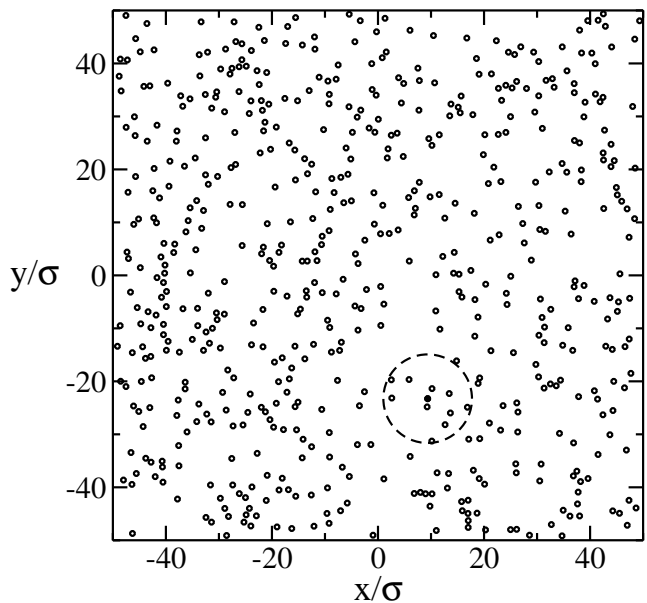}
\end{center}
%\vspace*{-2cm}
\caption{Snapshot of the system with $M=10m$. The intruder is in the
  center of the slashed circle and is
  represented by a
  filled symbol. }\label{foto6000}
\end{figure}

We have also investigated the behavior of the system for $M>M_c$ for
two different values of the intruder mass, $M=10m$ and $M=12m$. We
have performed MD simulations, finding that a stationary state is
reached in the long time limit. The obtained values for the stationary
partial temperatures are $X_s=1.3\pm 0.4$ and $\gamma=570\pm 260$
for $M=10m$ and $X_s=1.9\pm 0.5$ and $\gamma=800\pm 270$ for
$M=12m$. This strong non-equipartition is remarkable as the vertical
temperature is nearly three orders of magnitude larger than the
horizontal temperature. The measured kurtosis are 
$a_{2,xy}=0.1\pm 0.3$ and $a_{2,z}=-0.48\pm 0.09$ for $M=10$ and  
$a_{2,xy}=0.1\pm 0.3$ and $a_{2,z}=-0.52\pm 0.03$ for $M=12$, so that
the $z$-marginal velocity distribution is strongly non-gaussian. In
Fig. \ref{fzvzM10Fig}, $f_z$ is plotted as a function of
$\frac{v_z}{v_0}$ for $M=10m$, where a bimodal shape is clearly
observed. Similar results are obtained for $M=12m$. It must be
remarked here that the bath velocity distribution function is actually
disturbed for the analyzed values of the masses in the $M>M_c$
case. In effect, the bath partial temperatures deviate from the 
case without the presence of the intruder and the distribution
function deviates from the gaussian shape. Concretely, the
kurtosis of the bath in the $xy$ direction is 
$0.3\pm 0.12$ and $0.35\pm 0.12$ for $M=10m$ and $M=12m$
respectively. In the $z$-direction it is $0.20\pm 0.08$ and 
$0.24\pm 0.07$ for $M=10m$ and $M=12m$ respectively. In any case, it
is expected that, as increasing the number of bath particles, the
influence of the intruder in the bath can be minimized. 
Nevertheless, even in these extreme conditions where the bath is
highly disturbed by the intruder, the bath is still spatially
homogeneous as can be seen in Fig. \ref{foto6000}. 

Finally, we have also performed MD simulation in
the mass range $6<\frac{M}{m}<10$, but no clear conclusions can be
extracted from them, even the existence of a stationary state. The
closer to the critical mass, the larger the relaxation time to reach
the stationary state is, and more expensive simulations are needed to
study the behavior of the system with the same degree of accuracy as
in the $\frac{M}{m}\le 6$ and $\frac{M}{m}\ge 10$ cases.

\section{Conclusions and outlook}\label{secC}

In this paper we have analyzed the dynamics of an intruder, an
inelastic hard sphere, immersed in
a bath composed of inelastic hard spheres of 
the same diameter but different mass. The system is confined between
two hard parallel 
plates perpendicular to the vertical direction that inject energy into
the system in the direction perpendicular to them. A \emph{critical}
intruder mass, $M_c$, is identified for which the vertical temperature 
diverges when approaching it from below, remaining the horizontal
temperature finite. 
%If the intruder mass is not ``too close'' to the
%critical mass, the velocity distribution function of the intruder is
%approximately gaussian for $M<M_c$ and strongly non-gaussian for
%$M>M_c$, being the partial distribution function in the vertical
%direction a bimodal distribution. 
The mechanism triggering the
transition is identified in the context of a very simple model based
on the equations for the horizontal and vertical temperatures that are
derived from a kinetic theory description under clear and controlled approximations. 

In the theoretical study, it is assumed that the bath is in the
corresponding spatially homogeneous stationary state and that it is not disturbed by
the presence of the intruder. The dynamics of the distribution
function of the intruder is given by a Boltzmann-Lorentz-like equation
with two kind of collisional terms: one that takes into account the
collisions between the intruder and the bath particles and another that
takes into account the collisions between the intruder with the hard
walls. The former is modified with respect to the non-confined case in
order to take into account that
only the collisions compatible with the constraints are possible. The
kinetic equation is solved for spatially homogeneous states assuming
that the distribution function is a two-temperatures gaussian
corresponding to the vertical and horizontal temperatures. Under these
hypothesis, closed evolution equations for the partial temperatures are
obtained. Both equations contain a term  that comes from collisions
between the intruder and the bath particles that dissipates/injects
energy. The energy injection term that takes into account the collisions of the intruder with
the walls only appears in the vertical temperature equation, 
consistently with the fact that the walls inject energy in the
vertical direction. The fact that the collision between the particles
term is mass-dependent, while the intruder-wall term is
mass-independent, makes possible a stationary state only if the
intruder mass is smaller that certain \emph{critical} mass,
$M_c$. If $M>M_c$, there is not stationary state in the gaussian
approximation and the vertical temperature diverges in the long-time
limit.  

A very good agreement between the MD 
simulation results and the theoretical predictions is obtained for
$M<M_c$ both, for the dynamics and the stationary values reached in
the long-time limit. MD simulations show that the intruder velocity
distribution function is, in effect, close to a two-temperatures
gaussian if $M<M_c$ and the mass is not too close to the critical
mass while, close to the critical mass, the distribution function is not
gaussian anymore. 
Above the critical mass, the simulation results show that a stationary
state is reached but with a vertical temperature orders of magnitude
larger than the horizontal temperature and being the partial distribution
function in the vertical direction a 
bimodal distribution. 
Moreover, MD results also show that the bath is
not disturbed by the intruder if $M<M_c$ and $M$ is not close to the
critical value, consistently with the theoretical analysis. For $M$
close to the critical mass or $M>M_c$, the distribution function is
strongly disturbed and the gaussian approximation fails. Physically, the reason
is that, as the vertical temperature of the intruder is so large, there
can be collisions between a particle of the bath and the intruder
having a extremely high vertical velocity that affect the dynamics of
the bath. 

The present study opens the possibility of further studies that
are under investigation. First, the problem of diffusion. It seems
that, in the region where the intruder distribution function is
gaussian, we should have normal diffusion. When the distribution
function is not gaussian, the situation is not clear. In any case, what it
is clear is that the non gaussianities will modify the transport
coefficients even if the diffusion is still normal. Second, the
microscopic origin of the bimodal distribution. It seems plausible to
tackle the problem for very large masses by studying the corresponding
Fokker-Planck equation. Finally, taking into account that the
microscopic origin of the instability is very general and simple, we
think that many of the features studied in the paper could be observed in
actual experiments. Although a quantitative  agreement
  of the results reported here with experiments is not to be
  expected, due to the several simplifications introduced in the
  theoretical model, e.g.  
neglecting friction and rotation of the particles, a qualitative
agreement looks quite possible, since the elements considered in our
description are also present in experiments. In particular, the
existence of a \emph{critical} mass for which the vertical 
temperature diverges when approaching it from below, and the transition
from the gaussian distribution to the bimodal distribution above the
critical mass.

\begin{acknowledgments}
This research was supported by Consejer\'ia de Econom\'ia,
Conocimiento, Empresas y Universidad de la Junta de Andaluc\'ia
(Spain) through Grant. US-1380729 and by the Ministerio de Ciencia e
Innovaci\'on (Spain) through Grant PID2021-126348NB-100 (both partially financed
by FEDER funds). 
\end{acknowledgments}

\appendix

\section{Velocity moments of the collisional term}\label{velMomen}
The objective of this Appendix is the evaluation of the function
$\mathcal{G}$ defined in Eq. (\ref{ecG}). 
The evaluation of $\mathcal{H}$ given by Eq. (\ref{ecH}) follows
similar lines and will not be given. By standard arguments,
the expression of $\mathcal{G}$ given by Eq. (\ref{mathGeq}) can be rewritten as 
\begin{eqnarray}
\mathcal{G}=\frac{\sigma^2}{2(h-\sigma)n}\int d\mathbf{v}\int
d\mathbf{v}_1\int_{\sigma/2}^{h-\sigma/2} dz\int_{\Sigma(z)} d\sig
f_1(\mathbf{v}_1)f(\mathbf{v},t)\nonumber\\
\abs{\mathbf{g}\cdot\sig}(b_{\sig}-1)(v_x^2+v_y^2). \nonumber\\
\end{eqnarray}
By using the collision rule, Eqs. (\ref{colRule1}) and
(\ref{colRule2}), it is 
\begin{eqnarray}
(b_{\sig}-1)(v_x^2+v_y^2)=\left(\frac{m}{m+M}\right)^2(1+\alpha)^2
(\mathbf{g}\cdot\sig)^2(\hat{\sigma}_x^2 +\hat{\sigma}_y^2)
\nonumber\\+\frac{2m}{m+M}(1+\alpha)
(\mathbf{g}\cdot\sig)(v_x\hat{\sigma}_x+v_y\hat{\sigma_y}), \nonumber\\
\end{eqnarray}
and $\mathcal{G}$ can be expressed as
\begin{eqnarray}\label{mathcalGApp}
&&\mathcal{G}=\frac{\sigma^2}{2(h-\sigma)n}\left\{\left(\frac{m}{m+M}\right)^2(1+\alpha)^2
\right.\nonumber\\
&&\int_{\sigma/2}^{h-\sigma/2}dz\int_{\Sigma(z)}d\sig
G_1(\sig)(\hat{\sigma}_x^2+\hat{\sigma}_y^2)\nonumber\\
&&\left.+\frac{2m}{m+M}(1+\alpha)
  \int_{\sigma/2}^{h-\sigma/2}dz\int_{\Sigma(z)} d\sig G_2(\sig)
\right\}, 
\end{eqnarray}
where we have introduced
\begin{eqnarray}
G_1(\sig)=\int d\mathbf{v}\int
  d\mathbf{v}_1f_1(\mathbf{v}_1)f(\mathbf{v},t)
\abs{\mathbf{g}\cdot\sig}^3, \nonumber\\
\\
G_2(\sig)=\int d\mathbf{v}\int
  d\mathbf{v}_1f_1(\mathbf{v}_1)f(\mathbf{v},t)
\abs{\mathbf{g}\cdot\sig}(\mathbf{g}\cdot\sig)(v_x\hat{\sigma}_x+v_y\hat{\sigma}_y). 
\nonumber\\
\end{eqnarray}
To evaluate the above integrals, it is convenient to use the following variables 
\begin{eqnarray}
\mathbf{c}&=&\frac{1}{\Omega}(v_x\mathbf{e}_x+v_y\mathbf{e}_y)
+\frac{1}{\Omega_z}v_z\mathbf{e}_z, \\
\mathbf{c}_1&=&\frac{1}{w}(v_{1 x}\mathbf{e}_x+v_{1 y}\mathbf{e}_y)
+\frac{1}{w_z}v_{1 z}\mathbf{e}_z. 
\end{eqnarray}
Taking into account the Gaussian character of $f_1$ and $f$ (see
Eqs. (\ref{f1maintext}) and (\ref{dfGaussian})), it is obtained
\begin{eqnarray}
G_1(\sig)&=&\frac{n_1n}{\pi^3}\int
d\mathbf{X}e^{-X^2}\abs{\mathbf{X}\cdot\mathbf{u}_a}^3
a^3, \\
G_2(\sig)&=&\frac{n_1n}{\pi^3}\int
d\mathbf{X}e^{-X^2}\abs{\mathbf{X}\cdot\mathbf{u}_a}(\mathbf{X}\cdot\mathbf{u}_a)a^2
(c_x\hat{\sigma}_x+c_y\hat{\sigma}_y), \nonumber\\
\end{eqnarray}
where the time dependence in $G_1$ and $G_2$ has not been explicitly
written because it comes entirely through the thermal velocities
$\Omega$ and $\Omega_z$. We have also introduced the six-dimensional variable
$\mathbf{X}=(c_{1 x}, c_{1 y}, c_{1 z}, c_x, c_y, c_z)$, the vector
$\mathbf{a}=(w\hat{\sigma}_x, w\hat{\sigma_y}, w_z\hat{\sigma}_z, 
-\Omega\hat{\sigma}_x, -\Omega\hat{\sigma_y},
-\Omega_z\hat{\sigma}_z)$, its modulus $a\equiv\abs{\mathbf{a}}$ and the unit vector
$\mathbf{u}_a\equiv\mathbf{a}/a$. Performing the gaussian integrals
and taking into account that
$a=[(w^2+\Omega^2)(\hat{\sigma}_x^2+\hat{\sigma}_y^2)
+(w_z^2+\Omega_z^2)\hat{\sigma}_z^2]^{1/2}$, it is obtained
\begin{eqnarray}
G_1(\sig)&=&\frac{n_1n}{\sqrt{\pi}}[(w^2+\Omega^2)(\hat{\sigma}_x^2+\hat{\sigma}_y^2)
+(w_z^2+\Omega_z^2)\hat{\sigma}_z^2]^{3/2}, \\
G_2(\sig)&=&-\frac{n_1n}{\sqrt{\pi}}\Omega^2
[(w^2+\Omega^2)(\hat{\sigma}_x^2+\hat{\sigma}_y^2)
+(w_z^2+\Omega_z^2)\hat{\sigma}_z^2]^{1/2}(\hat{\sigma}_x^2+\hat{\sigma}_y^2). 
\nonumber\\
\end{eqnarray}
To obtain the desired expression for $\mathcal{G}$, the above
functions have to be inserted in Eq. (\ref{mathcalGApp}). Then, the
relevant integrals to be performed are
$\int_{\sigma/2}^{h-\sigma/2}dz\int_{\Sigma(z)} d\sig G_1(\sig) 
(\hat{\sigma}_x^2+\hat{\sigma}_y^2)$ and 
$\int_{\sigma/2}^{h-\sigma/2}dz\int_{\Sigma(z)} d\sig G_2(\sig)$. 
Performing the angular integration and introducing the
dimensionless variables $y\equiv\frac{z-\frac{\sigma}{2}}{\sigma}$, 
the following result is obtained
\begin{eqnarray}
\int_{\sigma/2}^{h-\sigma/2}dz\int_{\Sigma(z)} d\sig G_1(\sig) 
(\hat{\sigma}_x^2+\hat{\sigma}_y^2) 
=4\sqrt{\pi}n_1n\sigma
\int_0^{\epsilon}dy(\epsilon-y) \nonumber\\ 
(1-y^2)[(w^2+\Omega^2)(1-y^2)
+(w_z^2+\Omega_z^2)y^2]^{3/2}, \nonumber\\ 
\\
\int_{\sigma/2}^{h-\sigma/2}dz\int_{\Sigma(z)} d\sig G_2(\sig) 
=-4\sqrt{\pi}n_1n\sigma\Omega^2\int_0^{\epsilon}dy(\epsilon-y)\nonumber\\
(1-y^2) [(w^2+\Omega^2)(1-y^2)
+(w_z^2+\Omega_z^2)y^2]^{1/2}. \nonumber\\
\end{eqnarray}
By inserting the above expressions in Eq. (\ref{mathcalGApp}), the
expression of the main text, Eq. (\ref{ecG}), is obtained.

\end{document}